\UseRawInputEncoding
\documentclass[preprint,12pt]{elsarticle}
\usepackage{amssymb}
\usepackage{placeins}
\usepackage{rotating}
\usepackage{colortbl}
\usepackage{lscape}
\usepackage{longtable}
\usepackage{adjustbox}
\usepackage{pdflscape}
\newcolumntype{L}{>{\centering\arraybackslash}m{6cm}}
\newcolumntype{M}{>{\centering\arraybackslash}m{2.5cm}l}

\biboptions{sort&compress}

\journal{Biomedical Signal Processing and Control}

\begin{document}

\begin{frontmatter}

\title{A Statistical Approach for Synthetic EEG Data Generation}
\author[inst1]{Gideon Vos}

\affiliation[inst1]{organization={College of Science and Engineering, James Cook University},
            addressline={James Cook Dr}, 
            city={Townsville},
            postcode={4811}, 
            state={QLD},
            country={Australia}}

\author[inst1]{Maryam Ebrahimpour}
\author[inst2]{Liza van Eijk}
\author[inst3]{Zoltan Sarnyai}
\author[inst1]{Mostafa Rahimi Azghadi}

\affiliation[inst2]{organization={College of Health Care Sciences, James Cook University},
            addressline={James Cook Dr}, 
            city={Townsville},
            postcode={4811}, 
            state={QLD},
            country={Australia}}

\affiliation[inst3]{organization={College of Public Health, Medical, and Vet Sciences, James Cook University},
            addressline={James Cook Dr}, 
            city={Townsville},
            postcode={4811}, 
            state={QLD},
            country={Australia}}
\begin{abstract}
\paragraph{Introduction}
\noindent Electroencephalogram data plays a critical role in understanding and diagnosing neurological conditions. However, recording EEG data is both costly and time-consuming, particularly when aiming to build large datasets required for training machine learning models. This limitation has driven interest in synthetic data generation as a means to augment existing datasets. Synthetic data not only reduces the dependency on extensive real-world recordings but also accelerates research by providing readily available training samples. While prior research has examined diverse approaches to data augmentation, the generation of high-fidelity synthetic EEG data that faithfully retains the nuanced signatures of emotional and cognitive states remains an open challenge. This study introduces a novel, statistically grounded framework that leverages correlation structure analysis and probabilistic sampling to produce synthetic EEG data with preserved signal integrity and enhanced applicability for emotion-related neuroinformatics.

\paragraph{Methods}
\noindent Correlation analysis was used to determine interdependencies between frequency bands in an original EEG dataset. Next, synthetic EEG samples were generated by leveraging random sampling techniques, guided by the correlation structure of the original EEG data. Synthetic samples were next tested against the original dataset using correlation analysis to ensure fidelity, and samples with high correlation were retained. Finally, the generated synthetic EEG data was subjected to distribution analysis and machine learning classification models trained to distinguish between original and synthetic samples, serving as a benchmark for the quality of the synthetic data. To support reproducibility and progress, all code from this study is publicly available on GitHub at https://github.com/xalentis/SyntheticEEG.

\paragraph{Results}
\noindent The synthetic EEG data generated using our proposed method exhibits high fidelity to the original dataset, while preserving subject emotional and mental health state. Similar correlation coefficients between the synthetic and original data confirmed the preservation of the underlying structure, with the synthetic EEG data matching the distribution of the original EEG data, while PERMANOVA analysis showing no statistical difference. A Random Forest machine learning classification model trained to classify synthetic versus original samples performed no better than random guessing, indicating the inability to distinguish between the two datasets.

\paragraph{Conclusion}
\noindent This study presents a novel, statistically driven approach for generating synthetic EEG data tailored to brain health research. Unlike complex generative models such as Generative Adversarial Networks (GANs) and Variational Autoencoders (VAEs), our method combines correlation structure analysis with randomized sampling to create data that closely mirrors real EEG patterns. The indistinguishability of synthetic samples from real ones, as demonstrated by classification models, underscores the approach’s fidelity and robustness. Our proposed technique offers a computationally lightweight, interpretable, and scalable alternative that lowers barriers to dataset expansion while preserving patient privacy. This streamlined approach has the potential to accelerate innovation in EEG-based machine learning, enabling scalable applications across diverse neurological and psychiatric research areas.

\end{abstract}


\begin{highlights}
\item EEG data can play a key role in diagnosing and monitoring brain-related conditions and applications, but collecting large datasets is resource and time-intensive.
\item Synthetic data offers a means to reduce reliance on extensive real-world recordings, accelerate research, and safeguard patient privacy.
\item Existing methods for generating synthetic EEG data often rely on advanced deep learning models that require substantial computational resources and specialized technologies.
\item The proposed method employs standard correlation analysis and random sampling to generate synthetic EEG data while maintaining the integrity of key signals. 
\item In favor of reproducible research and to advance the field, all programming code used in this study is made publicly available. 
\end{highlights}

\begin{keyword}
Machine Learning \sep EEG \sep Synthetic Data
\PACS 07.05.Mh \sep 87.19.La
\MSC 68T01 \sep 92-08
\end{keyword}

\end{frontmatter}



\section{Introduction}
\noindent Electroencephalography (EEG) is a non-invasive technique that records electrical activity in the brain through electrodes placed on the scalp. It has become a cornerstone in diagnosing and monitoring a variety of psychiatric \cite{Park2021, Newson2019} and neuropsychiatric disorders, including epilepsy \cite{Rasheed2021, Dash2025, Li2025}, depression \cite{Carrle2023, Xi2025, Boby2025} and schizophrenia \cite{RuizdeMiras2023, Ellis2022}. EEG is particularly valued for its ability to capture real-time brain activity, offering clinicians and researchers a window into the dynamic processes of the brain. Its affordability, portability, and non-invasive nature make it a practical tool in both clinical and research contexts. \\

\noindent A critical limitation in EEG-based research and clinical applications is the scarcity of high-quality, labeled datasets. The recording and collection of EEG data is time-consuming, resource-intensive \cite{Ney2024}, and subject to stringent privacy regulations \cite{Boudewyn2023, Arora2024}. Furthermore, differences in EEG hardware, sensor configurations, preprocessing pipelines and  variability in recording protocols \cite{Vos2024}, often lead to challenges in creating large, standardized datasets. These limitations hinder the training and validation of machine learning (ML) models, which require diverse and representative data to achieve robust performance. Addressing these challenges is essential for advancing the use of ML in EEG analysis, particularly in developing diagnostic tools and personalized treatment strategies.\\

\noindent Synthetic data generation has emerged as a promising solution to these challenges \cite{Rujas2025, Murtaza2023, Pantanowitz2024, Juwara2024, Pezoulas2024}. Synthetic data refers to artificially generated datasets that mimic the statistical properties and structural patterns of real data without compromising individual privacy. This approach, first conceptualized over three decades ago \cite{rubin1993statistical, little1993statistical}, has gained significant traction in recent years, driven by advancements in artificial intelligence (AI). Techniques such as generative adversarial networks (GANs) \cite{goodfellow2014gan, Carrle2023}, ensemble models \cite{Vos2023}, and latent diffusion models \cite{rombach2021high} have been employed to generate synthetic biomarker and EEG data that closely resembles real recordings. These synthetic datasets enable researchers to overcome data scarcity, enhance model generalizability, and facilitate data sharing across institutions without violating privacy regulations.\\

\noindent The potential applications of synthetic EEG data in ML are vast. By augmenting real datasets with synthetic data, researchers can address bias \cite{Juwara2024, Draghi2024}, improve the performance of predictive models \cite{HernandezMatamoros2020, Siddhad2024, Rasheed2021, Khosravi2024} and democratize research through the publication of open data \cite{Pezoulas2024, Kesteren2024}. Synthetic data also allows for the simulation of diverse patient populations, ensuring that ML models are trained on data that reflects the variability in real-world scenarios. This is particularly important in personalized medicine, where treatment recommendations must account for individual differences in brain activity and response patterns.\\

\noindent Despite its advantages, the use of synthetic data in medical research is not without challenges \cite{Rankin2020, Pantanowitz2024, Arora2024, Kesteren2024}. Ensuring the fidelity and reliability of synthetic EEG data is critical \cite{Vallevik2024}, as any discrepancies between synthetic and real data could impact the performance of ML models. Additionally, the ethical implications of synthetic data generation, including potential misuse and the need for transparency in data creation processes, must be carefully considered \cite{Arora2024}. Addressing these challenges requires ongoing research and the development of standardized guidelines for the generation and use of synthetic data in healthcare.\\

\section{Related Work}
\noindent Rujas \emph{et al.} \cite{Rujas2025} conducted a systematic review examining the application of synthetic data generation in healthcare, reporting that 36\% of studies focused on its use in developing image classification machine learning models. Similarly, Pezoulas \emph{et al.} \cite{Pezoulas2024} highlighted a significant rise in publications exploring synthetic data in healthcare. They identified cost and time efficiency as key drivers for synthetic data adoption, followed by the enhancement of privacy protections. Their review also revealed that deep learning-based synthetic data generators were employed in 72.6\% of studies, with statistical approaches accounting for 15.1\%. Among deep learning methods, GANs were the most frequently utilized. However, the synthetic generation of EEG data remains an under-explored area of research \cite{Rujas2025, Murtaza2023, Pantanowitz2024, Pezoulas2024, Carrle2023}\\

\noindent The use of deep learning techniques for synthetic data generation has been associated with a modest improvement in predictive accuracy, averaging 4\%, compared to models trained on original data \cite{Khosravi2024, Rasheed2021, Siddhad2024}. However, training these sophisticated models are computationally demanding and time-intensive \cite{Pezoulas2024}. Alternative methods including Variational Autoencoders (VAEs) \cite{Ozdenizci2021}, diffusion models \cite{Siddhad2024, Ozdenizci2021}, nonparametric tree-based techniques \cite{Rankin2020}, and Bayesian networks \cite{Rankin2020, Draghi2024}, have also been utilized in a number of prior studies with some success. Notably, Rankin \emph{et al.} \cite{Rankin2020} investigated the reliability of supervised machine learning models trained on synthetic data. Their findings indicated that tree-based classifiers, compared to deep-learning models, are particularly sensitive to synthetic data, with 92\% of models tested demonstrating reduced predictive accuracy, compared to those trained on original, non-synthetic data. \\

\noindent In this study, we present a computationally efficient and scalable method for generating synthetic EEG data using standard statistical techniques, such as random sampling and Spearman correlation analysis. In contrast to more complex generative models like GANs or VAEs, this approach is simpler to implement and requires significantly less computational overhead. It offers a practical and accessible solution for addressing data scarcity in EEG-based healthcare applications, while maintaining strong performance and supporting the development of more robust and privacy-preserving machine learning models.\\

\section{Methods}

\subsection{Datasets}
\noindent Three EEG datasets were utilized in this study (Table \ref{tab:datasets}). The EEG During Mental Arithmetic Tasks (Stress) dataset \cite{dataseteegstress} is provided pre-labeled for a relaxed state and an acute stress state during an arithmetic task. This dataset was additionally used to build a baseline stress prediction model using XGBoost \cite{XGBoost} with data split 70\%/30\% for training and testing, achieving an Area Under the Curve (AUC) of 1.0. The resulting model was then utilized to predict stress on the additional un-labelled datasets utilized in this study, with the aim of testing whether external emotion prediction can be reliably synthesized across datasets. \\

\noindent For the SAM40 dataset (SAM) \cite{datasetsam}, subjects were recorded for 25-second intervals while performing four different tasks: the stroop color-word test (SCWT), solving arithmetic questions, identification of symmetric mirror images, and a state of relaxation, with 3 trials recorded for each. The baseline stress prediction model was then applied to predict acute stress during each trial of the 4 tasks. \\

\noindent The third dataset, Mental Workload \cite{datasetworkload} was collected while subjects performed low, medium and high levels of two different complex tasks which included the N-back test game \cite{Kirchner1958} to enforce the short term memory, and a flight simulation. Performance scores were attributed during each task to measure each subject's ability to perform under each difficulty level. The baseline stress prediction model was again used to predict acute stress during each of the 3 difficulty levels for both tasks. For this dataset, the two tasks were separated into Mental Workload Dataset 1 and Mental Workload Dataset 2, resulting in a total of 4 experimental datasets for this paper. This separation was performed to evaluate data synthesis across tasks for the same subject group.

\begin{table}[!h]
\centering
\caption{\label{tab:datasets}Datasets utilized in this study.}
\resizebox{\textwidth}{!}{
\begin{tabular}{llcl}
\hline\hline
\textbf{Dataset}                            & \textbf{Categories}       & \textbf{Subjects}                                         \\
\hline
GoEmotions \cite{datasetworkload}                             & Emotiv Epoc Flex & 32                           & 40 (14F, 26M, mean age: 21.5 years)  \\
EEG During Mental Arithmetic Tasks \cite{dataseteegstress} & Neurocom         & 23                           & 35 (26F, 9M, mean age: 18.25 years)  \\
Mental Workload \cite{datasetworkload}                    & Emotiv Epoc X    & 14                           & 15 (age: 20-60)     \\                            
\hline
\end{tabular}
}
\end{table}
\FloatBarrier

\subsection{Pre-processing}
\begin{figure}[h!]
\centering
\includegraphics[height=0.6\textheight]{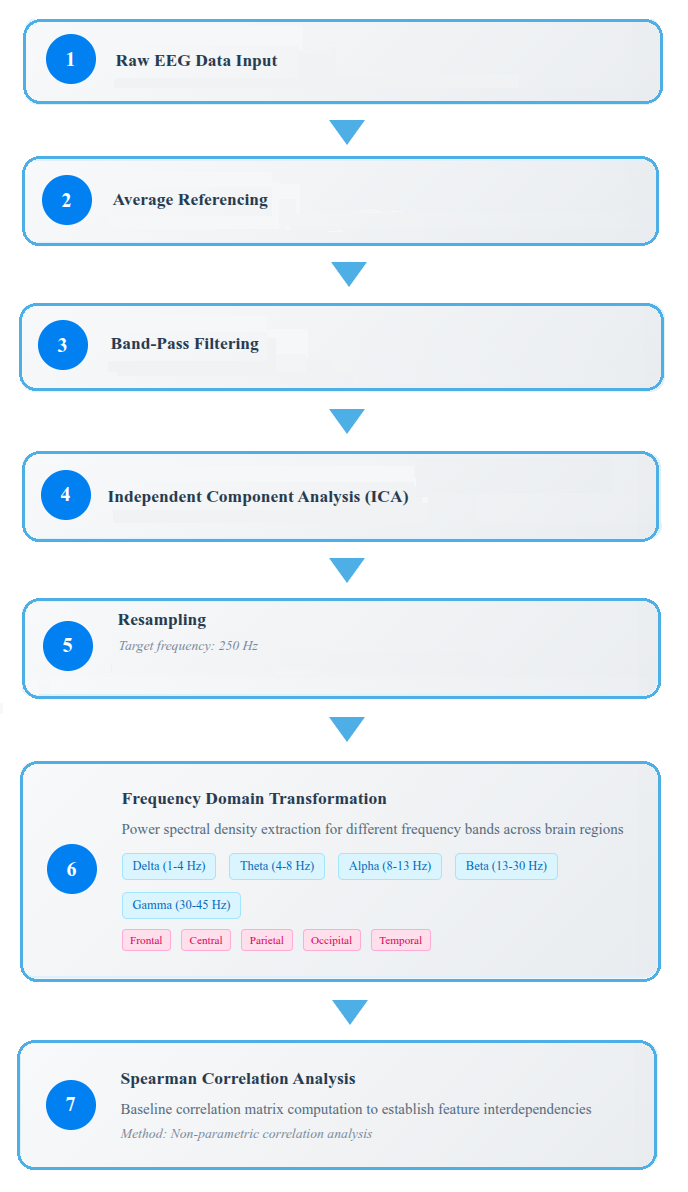}
\caption{\label{fig:figure1} EEG Pre-processing Pipeline. Steps 1 to 5 depict the uniform preprocessing applied to all datasets, including average referencing, band-pass filtering (1–45 Hz), ICA-based artefact removal using the MNE library, and resampling to 250 Hz. Step 6 shows transformation into the frequency domain to address potential sensor mismatches. Step 7 illustrates the computation of baseline Spearman correlations for subsequent synthetic EEG generation.}%
\end{figure}

\noindent To ensure a consistent and standardized pre-processing pipeline, artifact removal was uniformly applied to all four datasets (Figure \ref{fig:figure1}, steps 1 to 5), regardless of whether they were reported as artifact-free. Pre-processing steps included average referencing across all EEG electrodes, followed by the application of a band-pass filter with a frequency range of 1 Hz to 45 Hz. Independent Component Analysis (ICA) was subsequently performed using the MNE library \cite{GramfortEtAl2013a}, to identify and remove ocular, muscular and other potential artifacts. Finally, all datasets were resampled to 250 Hz to maintain frequency uniformity. \\

\noindent The artifact-free data of each dataset were then transformed into the frequency domain (Figure \ref{fig:figure1}, step 6). This step was designed to mitigate any potential impact of sensor mismatch arising from the use of different recording devices (see Table \ref{tab:datasets}) across the original 3 datasets. Finally, Spearman correlation analysis was conducted for each dataset, serving as a baseline threshold for the synthetic data generation phase (Figure \ref{fig:figure1}, step 7).

\FloatBarrier

\subsection{Synthesis}
\begin{figure}[h!]
\centering
\includegraphics[height=0.8\textheight]{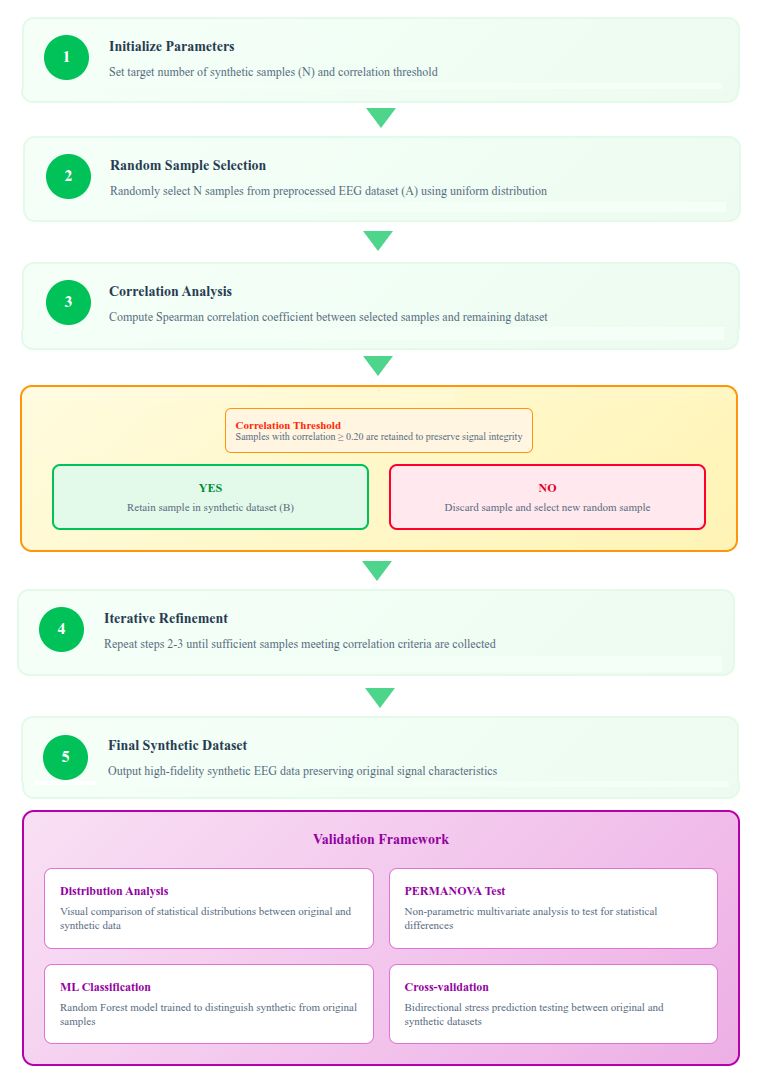}
\caption{\label{fig:figure2} Synthetic EEG Sample Generation Process. The pipeline begins with setting input parameters such as the number of desired synthetic samples \(N\) and the correlation threshold (step 1). Next, \(N\) random samples are selected from the pre-processed dataset (step 2), and Spearman correlation is calculated between these and the rest of the data (step 3). Samples below the threshold (0.20) are discarded. Steps 2 and 3 are iteratively repeated (step 4) until \(N\) sufficiently correlated samples are retained (step 5), forming the basis for generating realistic synthetic EEG data. The lower section of the figure outlines the subsequent validation framework applied to evaluate the quality of the generated samples.}%
\end{figure}

\noindent Generation of N synthetic samples involves the following step by step process (Figure \ref{fig:figure2}):

\begin{enumerate}
    \item N Random samples are selected from the pre-processed dataset (Figure \ref{fig:figure2}, step 2).
    \item Spearman correlation is performed between the selected N samples and the rest of the pre-processed data (Figure \ref{fig:figure2}, step 3).
    \item Samples exhibiting less correlation than a specified threshold (0.20 in the study experiments) are discarded, with the remaining retained.
    \item Steps 2 to 3 are repeated (Figure \ref{fig:figure2}, step 4) until sufficient samples are retained to equal the input parameter N (Figure \ref{fig:figure2}, step 5).
\end{enumerate}

\noindent In this study, 10-second epochs were generated for each dataset prior to conversion to the frequency domain (Figure \ref{fig:figure1}). Therefore, if 20 10-second epochs of synthetic data is required, N should be set to 20. Epoch durations of 10 seconds were selected due to the short recording length of subject samples within the original datasets; however, this is not mandatory and can be adjusted as required based on input data recording duration. 

\FloatBarrier

\subsection{Validation}\label{validation}
\noindent In order to validate the quality of the generated synthetic samples (refer Figure \ref{fig:figure2}), the following statistical and machine learning approaches were subsequently applied:

\begin{itemize}
    \item Distributions of original and synthetic samples were plotted and visually compared.
    \item A non-parametric PERMANOVA test was performed on the synthetic samples after testing for multivariate normality using the Shapiro-Wilk test.
    \item A Random Forest (RF) classification model was built after labeling and merging both original with synthetic samples, and trained to classify input data as either original (class 0) or synthetic (class 1).
    \item An RF model was trained on the original samples to predict the assigned stress label, and evaluated against the stress label of the synthetically generated samples.
    \item The prior step was repeated, by training on the synthetic samples to predict the stress label, and evaluated against the original data.
\end{itemize}

\noindent Synthetic samples constituting 70 10-second epochs were generated and evaluated for each of the four datasets (Table \ref{tab:datasets}) with the Mental Workload \cite{datasetworkload} dataset split into two datasets. This number (\textit{N=70}) of samples were selected due to the relatively small sample sizes of the original datasets utilized in this study. \\

\noindent In order to compare the proposed method to existing synthesis methods using GANs and VAEs, two experiments were performed using the EEG During Mental Arithmetic Tasks \cite{dataseteegstress} dataset. Distributions of original and synthetic samples were plotted and visually compared before training an RF model to predict the assigned stress label, and evaluated against the stress label of the synthetically generated samples.

\section{Results}

\begin{figure}[h!]
\centering
\includegraphics[width=\textwidth]{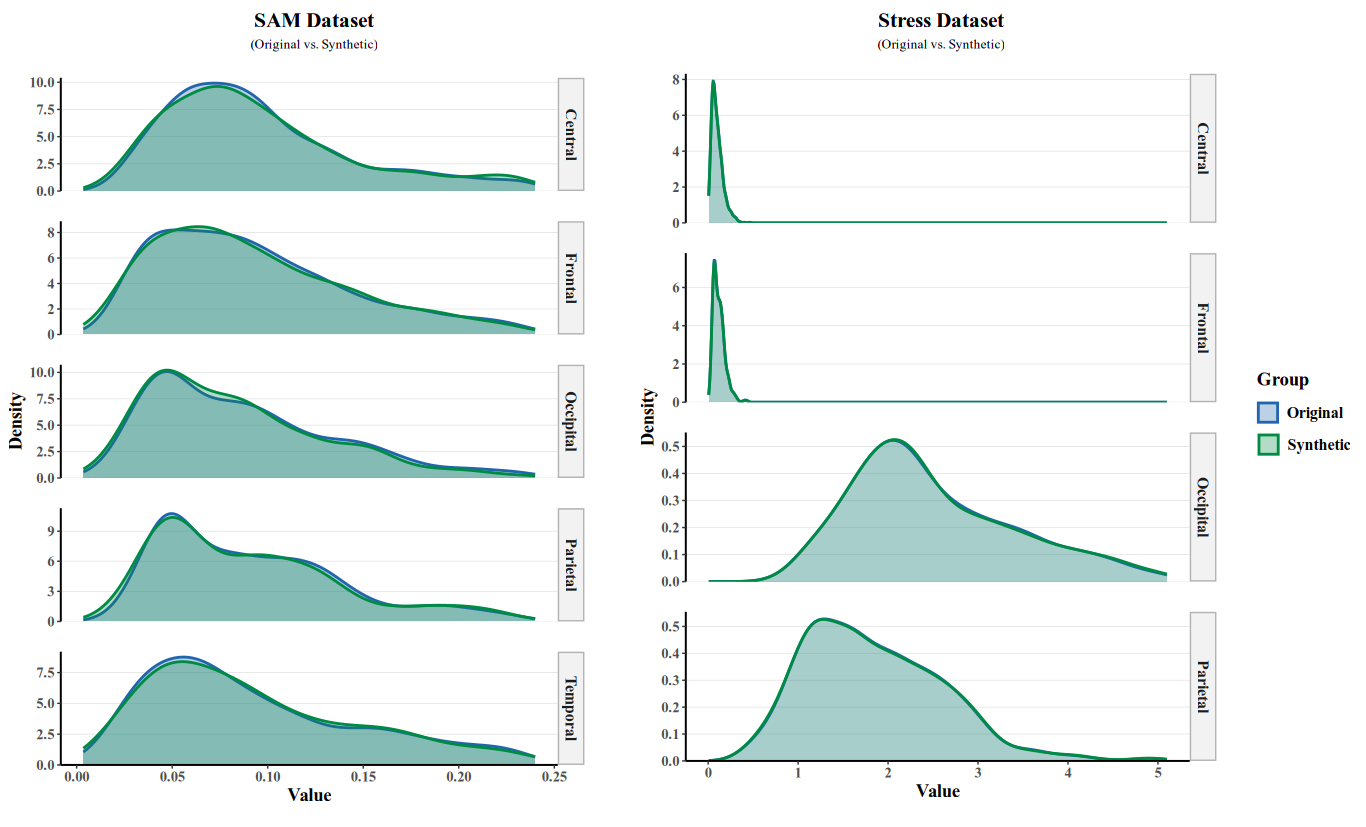}
\caption{\label{fig:figure3} Distribution Comparison of Original and Synthetic EEG Data for SAM and Stress Datasets. The figure shows a close resemblance between the statistical distribution of the synthetic and original data across five brain regions.}%
\end{figure}

\begin{figure}[h!]
\centering
\includegraphics[width=\textwidth]{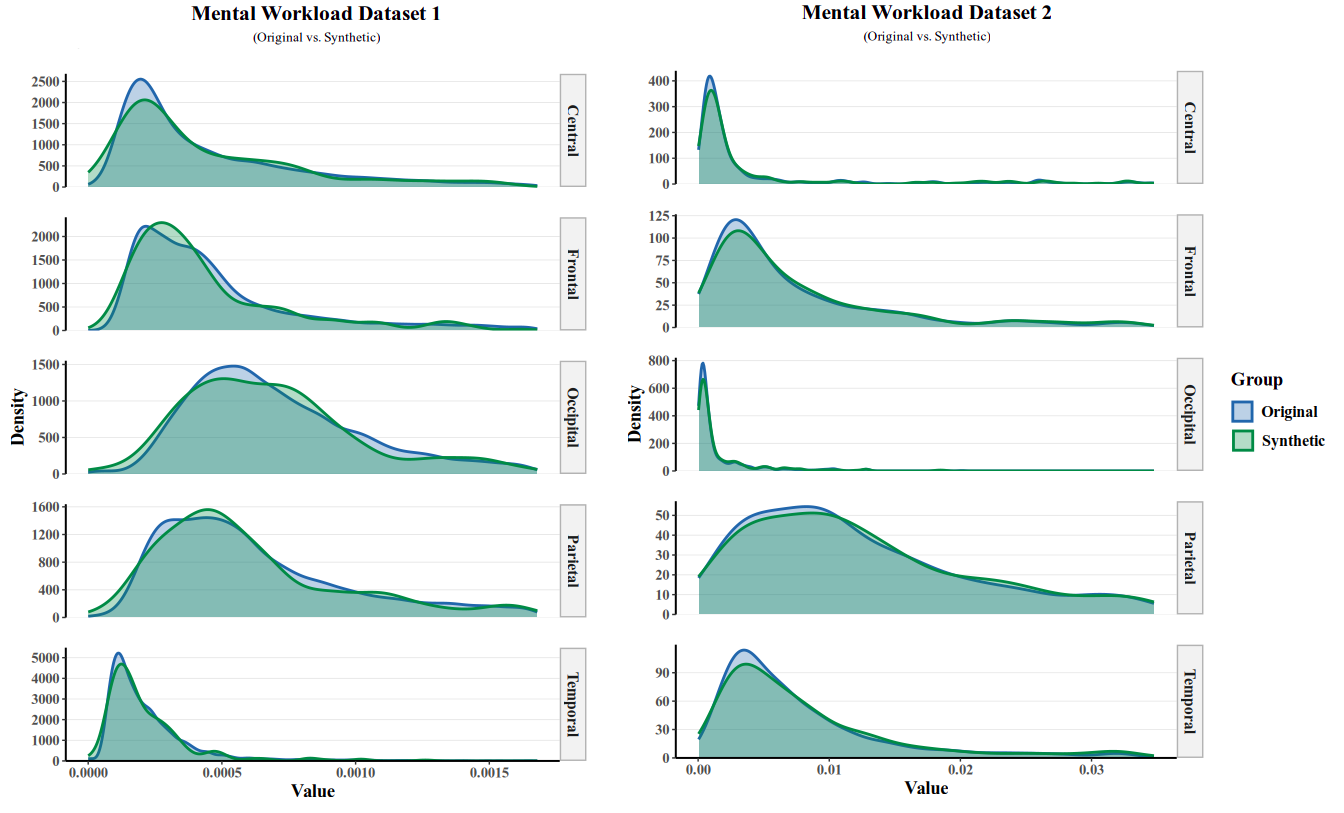}
\caption{\label{fig:figure4} Distribution Comparison of Original and Synthetic EEG Data for Mental Workload datasets 1 and 2. The figure shows a close resemblance between the statistical distribution of the synthetic and original data across five brain regions, with minor deviations observed in dataset 1.}%
\end{figure}

\noindent Following the validation process described in Section \ref{validation}, distribution plots for each dataset were generated, as shown in Figures \ref{fig:figure3} and \ref{fig:figure4}. Analyzing these distributions is a crucial step in evaluating the quality of synthetic EEG data \cite{Carrle2023}. By comparing the statistical properties of synthetic and real data, distribution analysis helps assess whether the synthetic samples capture the underlying patterns and variability of the original dataset. In this study, the distributions of the synthetic data for all four datasets closely resemble those of the original data, with minor deviations observed in Mental Workload Dataset 1 across four of the five recorded brain regions.

\FloatBarrier

\FloatBarrier

\begin{figure}[h!]
\centering
\includegraphics[width=\textwidth]{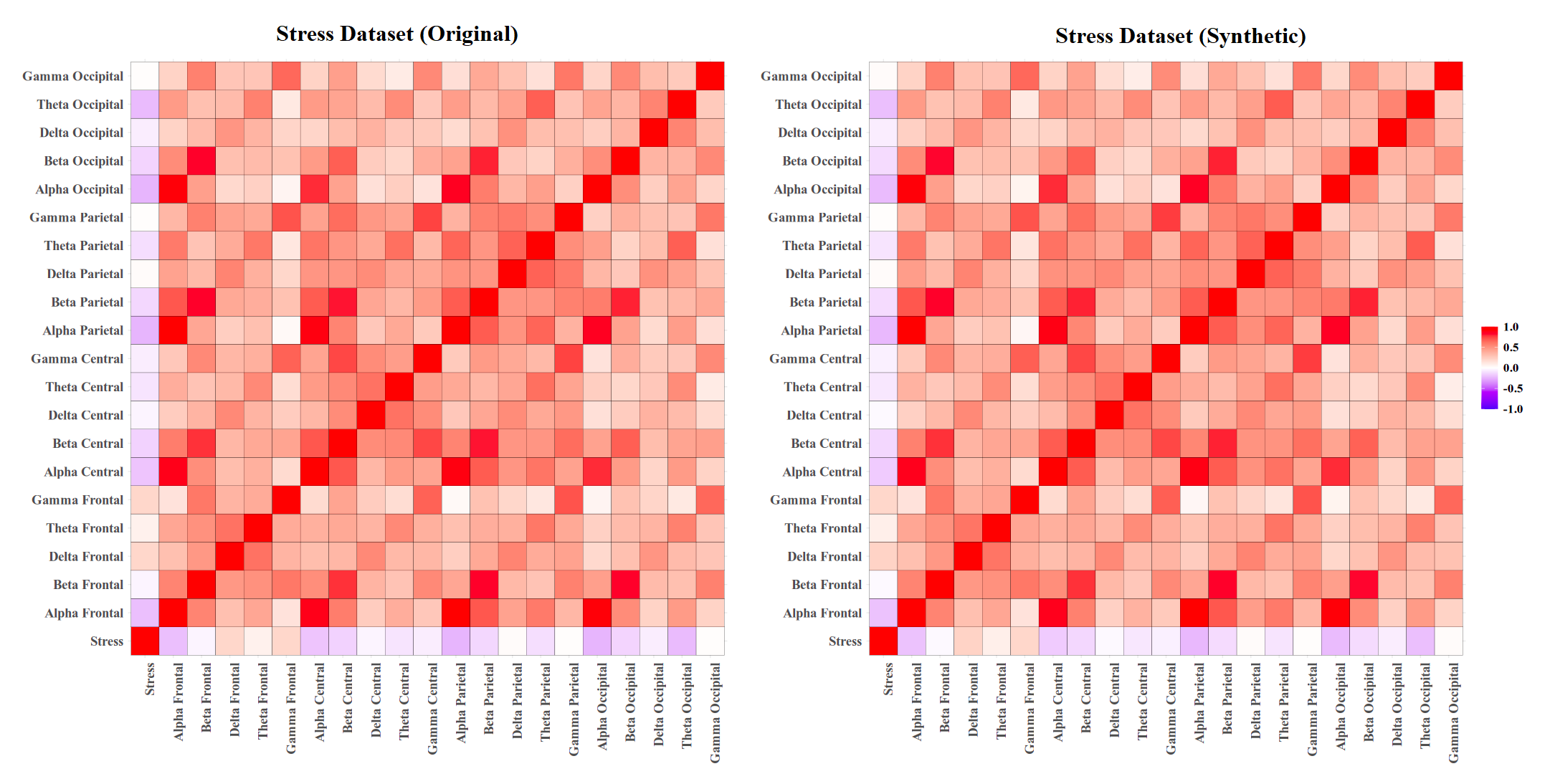}
\caption{\label{fig:figure5} Spearman correlation for Stress dataset showing strong preservation of inter-regional brain activity patterns between original (left) and synthetic (right) data, with particularly well-maintained correlations among occipital, parietal, central, and frontal regions.}
\end{figure}

\begin{figure}[h!]
\centering
\includegraphics[width=\textwidth]{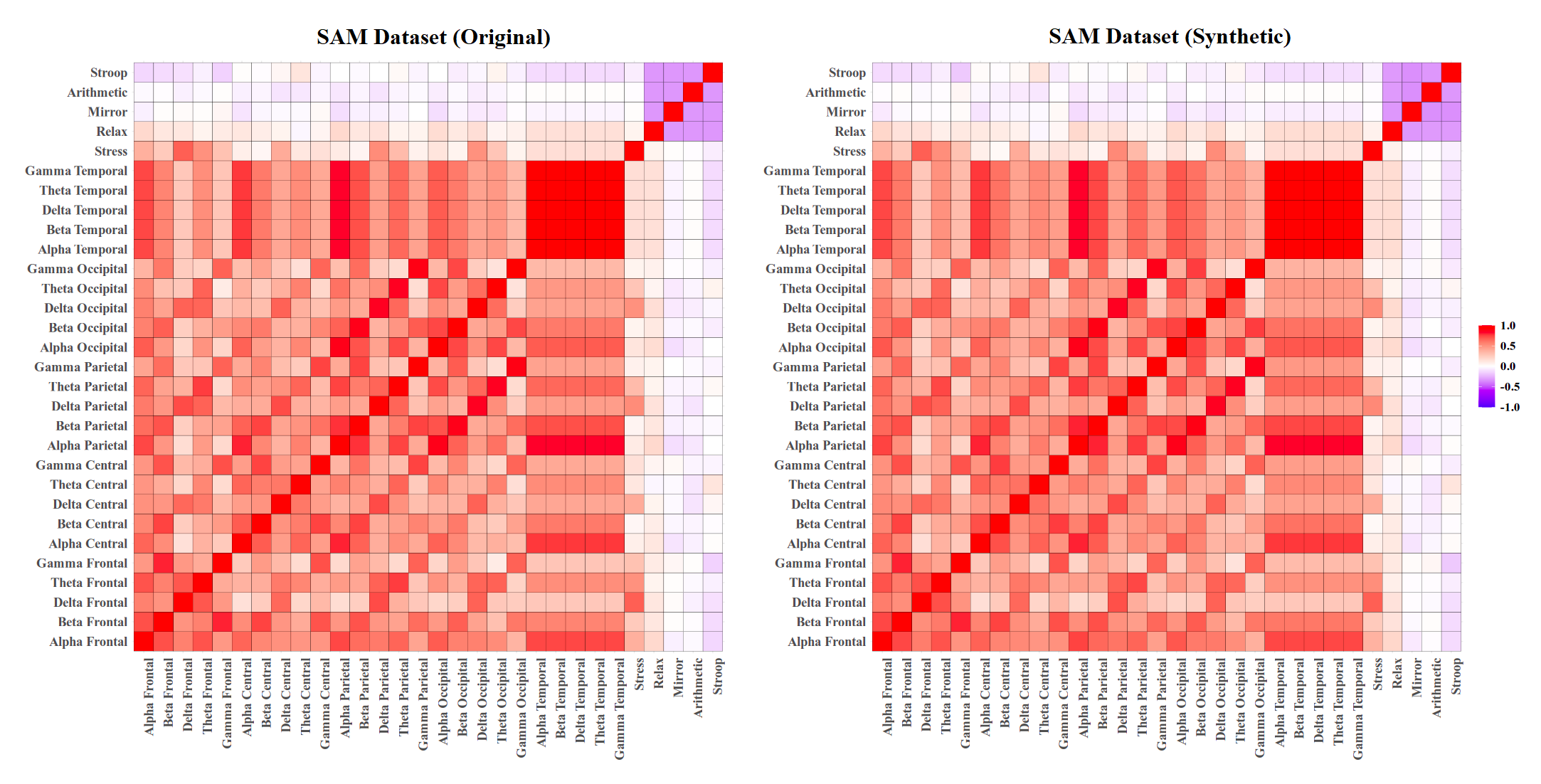}
\caption{\label{fig:figure6} SAM dataset correlation comparison demonstrating high fidelity synthetic data generation, with preserved strong positive correlations among temporal regions and maintained negative correlations between cognitive tasks (Stroop, Arithmetic, Mirror, Relax) and brain activity measures.}
\end{figure}

\noindent Spearman correlation analysis demonstrated consistent coefficients between the original and synthetic data across all four datasets. Notably, the stress labels in the Stress dataset are well-preserved in the synthetic data, as illustrated in Figure \ref{fig:figure5}. Figure \ref{fig:figure6} highlights the Spearman correlation between the original and synthetic versions of the SAM dataset, revealing minimal signal degradation for the arithmetic task, mostly in the gamma frontal and alpha central regions.\\

\noindent The acute stress measure, predicted using the XGBoost model trained on the original SAM dataset, is effectively preserved in the synthetic data, indicating strong signal fidelity. Notably, the relaxation state exhibits a significant negative correlation with the stress response, reinforcing the robustness of stress prediction. Region-specific activation patterns such as correlations in the frontal and temporal lobes during the Stroop test, and in the central and parietal regions during arithmetic tasks, are consistent with well-established neural correlates of cognitive load and stress processing \cite{Delorme2007, Hinterberger2003}. \\

\noindent The presence of stress-related signals in frontal, central, parietal, and occipital regions mirrors findings from real EEG data, where these regions are known to mediate emotional regulation, attention, and autonomic response \cite{Zhou2019, Andreassi2000}. The ability of the synthetic data to reproduce these spatial and spectral patterns underscores its neuro-physiological plausibility and practical value for downstream machine learning tasks that rely on region-specific EEG dynamics. \\

\FloatBarrier

\FloatBarrier

\begin{figure}[h!]
\centering
\includegraphics[width=\textwidth]{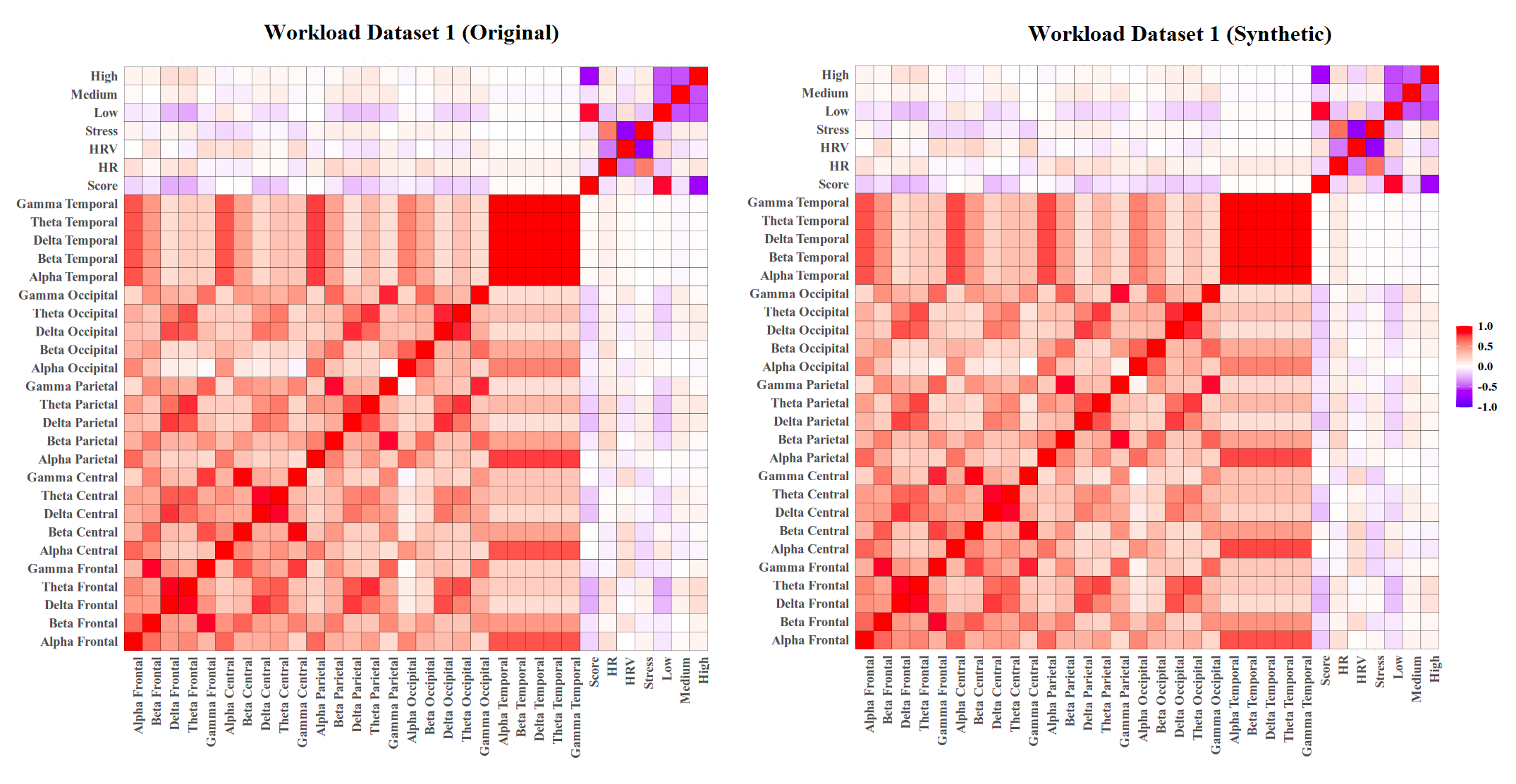}
\caption{\label{fig:figure7} Workload Dataset 1 correlation matrices revealing preservation of the original data structure in synthetic generation, including strong temporal region inter-correlations and appropriate negative correlations between workload measures (High, Medium, Low) and brain activity patterns.}
\end{figure}

\begin{figure}[h!]
\centering
\includegraphics[width=\textwidth]{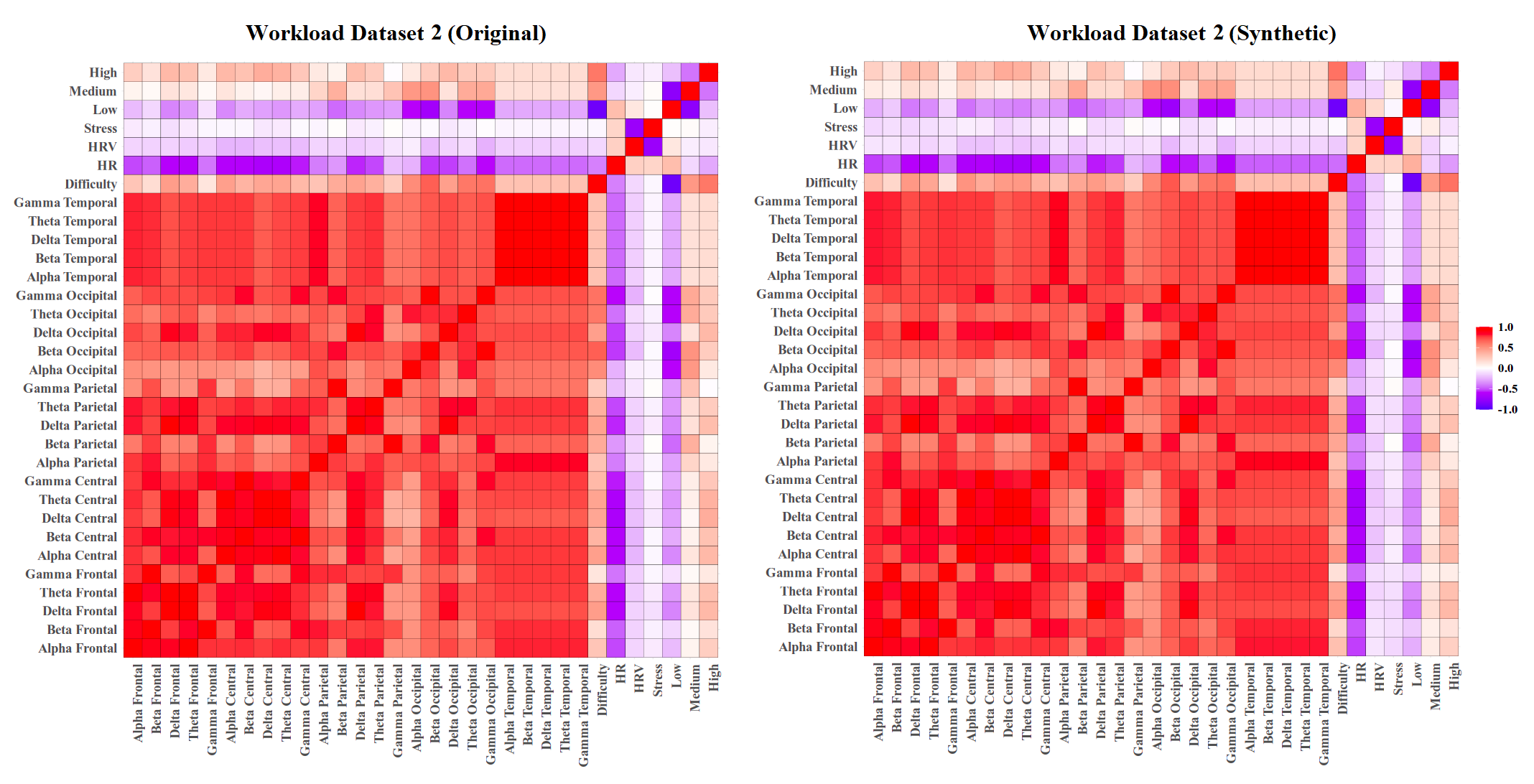}
\caption{\label{fig:figure8} Workload Dataset 2 correlation matrices showing robust synthetic data quality with preserved positive correlations among all brain regions and maintained negative relationships between difficulty/workload measures and neural activity, particularly in temporal areas.}
\end{figure}

\noindent Workload Datasets 1 and 2 exhibit stronger correlations in the synthesized data compared to the original data (Figures \ref{fig:figure7} and \ref{fig:figure8}). Additional non-EEG biomarkers present in the original datasets, such as heart rate (HR) and heart rate variability (HRV), were well-preserved in the synthetic data, alongside the stress measures predicted using the XGBoost model. Notably, Figures \ref{fig:figure7} and \ref{fig:figure8} reveal significant differences in correlation patterns between Workload Datasets 1 and 2, emphasizing the influence of varying experimental tests and protocols on EEG data.\\

\noindent In Workload Dataset 1, stress levels are generally low. However, when present, stress-related neural activity exhibits clear spatial and spectral organization. Notably, activity is localized in the frontal, central, and occipital regions which are commonly associated with cognitive control, sensorimotor integration, and visual processing \cite{Smith2009, Klimesch1999}. Beta and theta frequency bands further differentiate regional engagement, reflecting visual attention and fatigue, while theta activity in the frontal region aligns with working memory load \cite{Cavanagh2014, Harmony2013}. Task difficulty modulates these effects, with high-task conditions eliciting pronounced frontal and occipital correlations, whereas low-task conditions show inverse patterns. Medium difficulty engages temporal and central regions, highlighting dynamic shifts in neural resource allocation. The synthetic data preserves these region and frequency-specific patterns, reinforcing its validity for modeling real cognitive workload dynamics. \\

\noindent \noindent In contrast, Workload Dataset 2 elicits a substantially stronger stress response, reflected in widespread activation across the frontal, central, parietal, and occipital cortices. This broader engagement aligns with high cognitive load and distributed attention demands, consistent with prior studies of mental workload \cite{Roy2013, Berka2007}. Temporal lobe involvement, while slightly lower in correlation, is notable for its consistent activation across difficulty levels, likely reflecting language and sequential processing demands \cite{Pfurtscheller2001}. The synthetic data accurately replicated these spatial patterns and task-dependent modulations, supporting its application in downstream modeling of complex cognitive states. \\

\FloatBarrier

\FloatBarrier

\noindent PERMANOVA analyses were performed on both the SAM and Stress datasets to evaluate distributional differences between the original and synthetic samples. The results indicated no statistically significant differences, with p-values of \textit{p = 0.598} for the SAM dataset and \textit{p = 0.556} for the Stress dataset suggesting strong similarity between the two data sources. \\

\noindent Additionally, four RF classification models were trained on the original EEG data to predict stress in the synthetic counterparts, and vice versa. This approach served as a robust validation technique, since a well-trained machine learning model should be able to differentiate between two datasets if they contained distinct statistical properties or structural differences. If the model struggles to distinguish between original and synthetic data, it suggests that the synthetic data effectively replicates the key features of the real dataset. \\

\noindent In this study, the RF models exhibited near-random classification performance, with error rates of 47.62\% for the SAM dataset and 52.04\% for the Stress dataset, with values close to the 50\% mark showing indistinguishable distributions. These results indicate a high degree of similarity between the original and synthetic EEG data, reinforcing the validity of the synthetic samples for downstream machine learning applications. This approach provides an empirical, performance-based evaluation, complementing traditional statistical comparisons such as distribution analysis by offering strong evidence that the synthetic EEG data retains meaningful patterns and variability present in the original data. \\

\begin{figure}[h!]
\centering
\includegraphics[width=\textwidth]{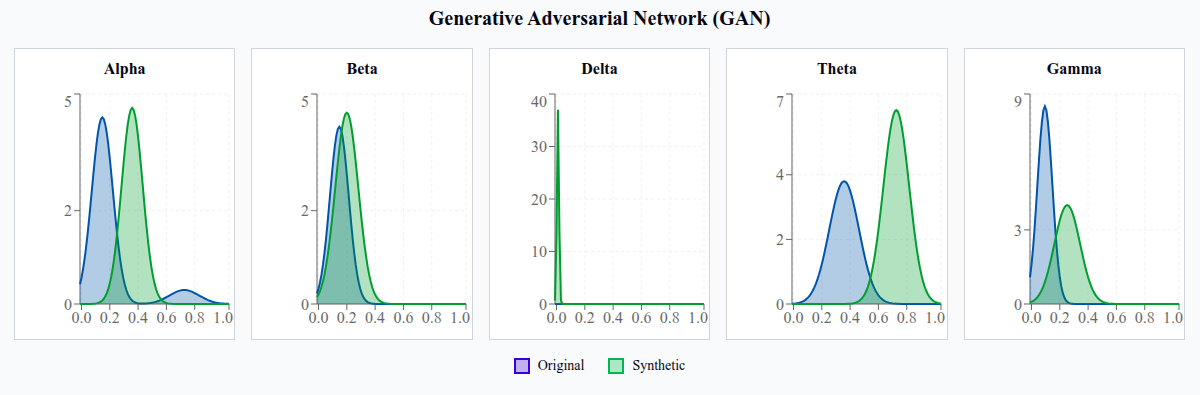}
\caption{\label{fig:figure9} Distribution analysis of original vs. synthetic data generated using a GAN. Substantial variation occurs across alpha, theta and gamma frequencies between the original and synthetically generated samples.}
\end{figure}

\begin{figure}[h!]
\centering
\includegraphics[width=\textwidth]{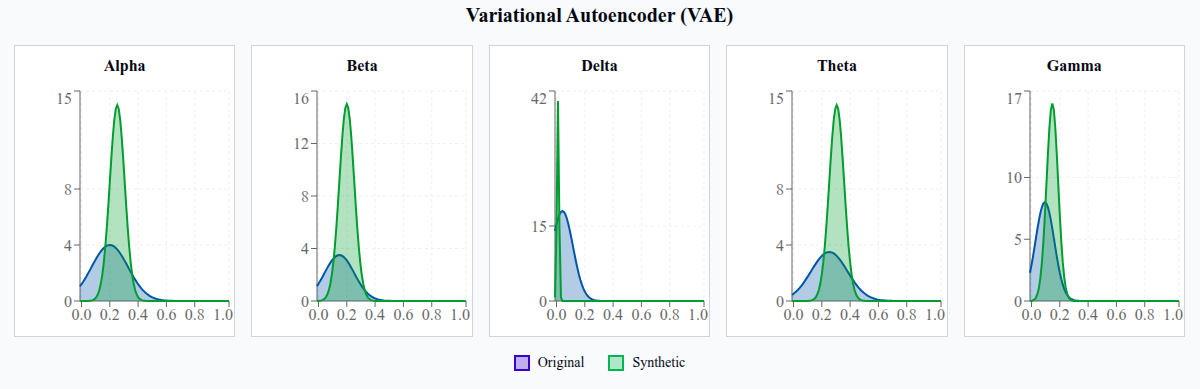}
\caption{\label{fig:figure10} Distribution analysis of original vs. synthetic data generated using a VAE. Substantial variation occurs across all frequencies between the original and synthetically generated samples.}
\end{figure}

\noindent Figures \ref{fig:figure9} and \ref{fig:figure10} show the distribution analysis when comparing the original EEG During Mental Arithmetic Tasks (Stress) data \cite{dataseteegstress} to its synthetic counterpart generated using a GAN (Figure \ref{fig:figure9}) and VAE (Figure \ref{fig:figure10}). The GAN consisted of a feedforward neural network with a 16-dimensional latent input mapped through a 64-unit hidden layer with ReLU activation, followed by a linear transformation to the 5-dimensional feature space (alpha, beta, delta, gamma, theta) using a Sigmoid output layer. The discriminator mirrored this structure, accepting the 5-dimensional input, processing it through a 64-unit hidden layer with ReLU, and outputting a scalar probability via a Sigmoid activation to distinguish real from synthetic samples.\\

\noindent This model was trained over 50 epochs using the Binary Cross-Entropy loss and the Adam optimizer with a learning rate of 0.001 for both networks. A batch size of 32 was employed during training. For the VAE, the encoder was mapped to the 5-dimensional input into a 64-unit hidden layer with ReLU activation, followed by two parallel linear layers that predict the latent mean and log-variance vectors, each of dimensionality 16. The decoder layer then reconstructs the input via a mirrored structure, projecting the latent vector through a 64-unit hidden layer, and ultimately to the original input space using a Sigmoid activation function to ensure output values remain within the [0, 1] range. This model was similarly trained for 50 epochs using the Adam optimizer with a learning rate of 0.001 and a batch size of 32. \\

\noindent Although GAN and VAE models represent state-of-the-art generative frameworks, the distributions of their synthetic outputs exhibit substantial variation and lack consistent fidelity to the original data. In contrast, the Random Forest (RF) model validation results (see section \ref{validation}) demonstrate that our statistical approach achieves comparable performance. This parity in performance underscores that the added complexity of GANs and VAEs does not translate into meaningful improvements in synthetic data quality for this particular application.

\FloatBarrier

\FloatBarrier

\section{Study Limitations}
\noindent The study's findings are limited by the use of a small number of EEG datasets, each with relatively modest sample sizes and a focus primarily on mental health and emotional states. While synthetic EEG data provides a viable alternative in scenarios with resource constraints, its current state does not fully replicate the complexity and authenticity of real EEG data, and remains insufficient as a complete substitute for original EEG recordings in research and clinical applications. Future work could explore the integration of additional physiological and contextual data to further enhance the utility and realism of synthetic datasets. Additionally, extending the validation framework to include domain-specific performance metrics and real-world applications, could provide deeper insights into the applicability of synthetic EEG data across diverse fields.

\section{Conclusion}
\noindent This study presents a novel, scalable, and cost-efficient method for generating synthetic EEG data that addresses key limitations of traditional EEG acquisition including acquisition cost and public data availability. By combining correlation analysis with random sampling, the approach produces synthetic EEG data that closely replicate the statistical and structural properties of real signals, including complex neuro-physiological relationships tied to emotion and stress. \\

\noindent Validation was performed using diverse datasets incorporating EEG alongside complementary biomarkers and emotion scores, with a rigorous five-step evaluation framework encompassing statistical tests and classifier-based assessments. This comprehensive validation process confirmed that the synthetic data preserved essential signal characteristics, variability, and meaningful physiological patterns relevant to brain function under stress and cognitive load. \\

\noindent Beyond technical advantages, synthetic EEG data generated via this approach substantially enhances patient privacy by eliminating direct links to individual subjects, thus mitigating ethical and security concerns associated with real EEG data. By delivering an accessible, validated, and privacy-preserving data generation solution, this work advances the field toward scalable and ethical EEG data augmentation, ultimately empowering broader research and machine learning applications across neurological and mental health domains. To encourage further exploration and application of this method, the complete source code used in this study is publicly available on GitHub at https://github.com/xalentis/SyntheticEEG.

 \bibliographystyle{elsarticle-num} 
 \bibliography{cas-refs}






\end{document}